\documentclass[aps,pre,twocolumn,showpacs,superscriptaddress]{revtex4-1}

\usepackage{setspace}
\usepackage{graphicx}
\usepackage{dcolumn}
\usepackage{bm}
\usepackage{amsmath,amssymb,mathtools}
\usepackage{color}
\definecolor{darkblue}{rgb}{0,0,0.6}
\definecolor{darkred}{rgb}{0.6,0,0}
\definecolor{darkgreen}{rgb}{0,0.6,0}

\usepackage[colorlinks=true,urlcolor=darkblue,citecolor=darkblue,linkcolor=darkred,hyperfootnotes=false]{hyperref}

\makeatother

\begin{document}

\preprint{APS/123-QED}

\title{Symmetry-breaking motility of penetrable objects in active fluids}

\author{Ki-Won Kim}
\affiliation{Department of Physics and Astronomy and Center for Theoretical Physics, Seoul National University, Seoul 08826, Republic of Korea}

\author{Yunsik Choe}
\affiliation{Department of Physics and Astronomy and Center for Theoretical Physics, Seoul National University, Seoul 08826, Republic of Korea}

\author{Yongjoo Baek}
\email{y.baek@snu.ac.kr}
\affiliation{Department of Physics and Astronomy and Center for Theoretical Physics, Seoul National University, Seoul 08826, Republic of Korea}

\begin{abstract}
We investigate how a symmetric penetrable object immersed in an active fluid becomes motile due to a negative drag acting in the direction of its velocity. While similar phenomena have been reported only for active fluids that possess polar or nematic order, we demonstrate that such motility can occur even in active fluids without any preexisting order. The emergence of object motility is characterized by both continuous and discontinuous transitions associated with the symmetry-breaking bifurcation of the object's steady-state velocity. Furthermore, we also discuss the relevance of the transitions to the nonmonotonic particle-size dependence of the object's diffusion coefficient.
\end{abstract}

\maketitle

\section{Introduction}
An active fluid is a fluid consisting of active particles, which utilize stored energy to propel themselves~\cite{RamaswamyARCMP2010,MarchettiRMP2013,RamaswamyJSM2017,JulicherRPP2018,GompperJPCM2020,BowickPRX2022}. Through current rectification, an asymmetric object immersed in an active fluid generally induces long-range density gradients~\cite{GalajdaJB2007,WanPRL2008,TailleurEPL2009,StenhammarSA2016} or persistent motion~\cite{AngelaniNJP2010,KaiserPRL2014,MalloryPRE2014,SmallenburgPRE2015}. These phenomena have been applied to the design of targeted delivery systems~\cite{KoumakisNComms2013} and self-starting micromotors~\cite{AngelaniPRL2009,SokolovPNAS2010,DiLeonardoPNAS2010}.

However, an asymmetric shape is not always necessary for such phenomena; there are various examples of symmetric objects that exhibit motility via symmetry breaking. Many of them feature preexisting order in the system. For instance, an active droplet with polar order in a passive fluid is known to develop splay instability, which in turn induces unidirectional motion~\cite{TjhungPNAS2012}. Conversely, a passive droplet inside a polar active gel can become motile by spontaneous creation of a topological defect~\cite{DeMagistrisSM2014}. Another possible scenario is when the object is highly flexible. Polymer chains in active fluids can spontaneously develop curvatures and turn into traveling structures~\cite{NikolaPRL2016,ShinPCCP2017,AporvariSoftMatter2020}. 

In this paper we show that neither an ordered medium nor a highly flexible object is needed for such motility to arise. Using a simple model of a symmetric penetrable object immersed in an active fluid lacking any order, we analytically describe the steady-state dynamics of the object. It turns out that the object motion by itself induces rectification, which creates a negative drag that acts in the direction of motion. While negative drag has been reported for transport by molecular motors~\cite{HowardARB2009} and contractile active nematics~\cite{FoffanoPRL2012}, it is also possible even in an ideal active gas, as recently discussed in \cite{GranekPRL2022}. While the study focused on the regime where the negative drag is so small that it affects only the diffusive properties of the object, here we investigate the case where the negative drag is strong enough, giving rise to persistent motion of the object via symmetry-breaking phase transitions.

Our results also have interesting implications for the nonmonotonic object-size dependence of effective diffusivity. While the phenomenon has been attributed to the interplay of diffusion and advection~\cite{KasyapPhysFluids2014,PattesonSoftMatter2016,BurkholderPRE2017,DyerPF2021,XuCJCP2021}, we discover that symmetry-breaking motility contributes an alternative mechanism.

The rest of this paper is organized as follows. In Sec.~\ref{sec:model} we introduce a simplified model of a symmetric penetrable object immersed in a one-dimensional (1D) active fluid. The drag force acting on this object is calculated in Sec.~\ref{sec:drag}, revealing the existence of the negative drag regime. In Sec.~\ref{sec:transitions} we use the mean-field theory to show that the negative drag gives rise to symmetry-breaking motility of the object via continuous and discontinuous phase transitions. The consequences of these transitions on the effective diffusivity of the object is discussed in Sec.~\ref{sec:diffusion}. Generalizations of the negative-drag mechanism to a broader range of systems, including the cases where the object-particle interactions are nonconservative and the system is two-dimensional (2D), are discussed in Sec.~\ref{sec:generalization}. We summarize our findings and contemplate possible future works in Sec.~\ref{sec:summary}.

\begin{figure*}
\includegraphics[width=\textwidth]{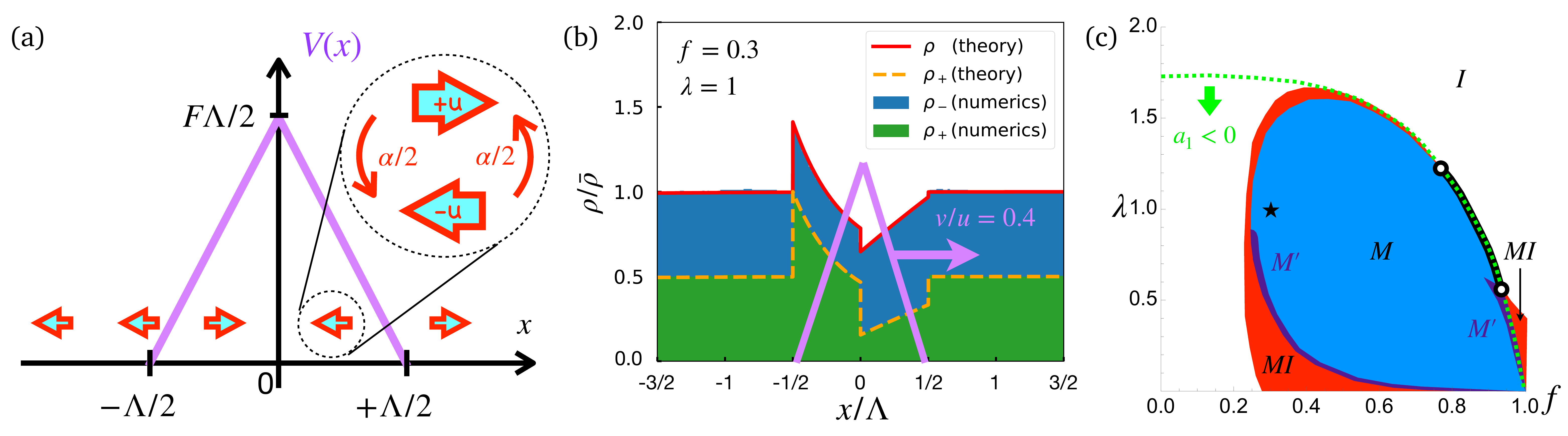}
\vspace{-0.25in}
\caption{\label{fig:main} (a) Schematic illustration of the model. (b) Density profile of the RTPs around a symmetric object moving to the right at a constant velocity. The numerics refer to the results of a particle-based simulation using $50\;000$ RTPs. The portions accounted for by the left-moving ($\rho_-$) and the right-moving ($\rho_+$) RTPs are distinguished using different shades. (c) Diagram showing the stable fixed points of the mean-field approximation in the large-$L$ limit. Negative drag is observed below the dotted line, and continuous transitions occur on the thick solid line. The star indicates the parameters used in (b).}
\end{figure*}

\section{Model} \label{sec:model}
We consider a symmetric, overdamped, penetrable object of size $\Lambda$ immersed in an active ideal gas on a 1D ring of length $L$ [see Fig.~\ref{fig:main}(a)]. The gas consists of $N$ run-and-tumble particles (RTPs) describing bacterial motion~\cite{SchnitzerPRE1993,BergBook2004}. Each RTP travels to the left or to the right at constant velocity $u$, flipping the direction at rate $\alpha/2$. The RTPs do not interact with each other but interact only with the object via the potential
\begin{align} \label{eq:V_def}
V(x) = \begin{cases}
 	F(x+\Lambda/2)	& \text{for $-\Lambda/2 \le x < 0$,} \\
 	-F(x-\Lambda/2)	& \text{for $0 \le x < \Lambda/2$,} \\
 	0			& \text{otherwise.}
 \end{cases}
\end{align}
Thus the RTP and the object repel each other at constant force $F$ whenever they overlap. We also assume that the thermal noise is negligible compared to the other forces. With these assumptions, each RTP obeys
\begin{align} \label{eq:RTP_motion}
    \dot{x}_i = -\mu\,V'(x_i-X) + u\,s_i(t)\quad\text{for $i \in \{1,\ldots,N\}$},
\end{align}
where $x_i$ is the position of the $i$-th RTP, $\mu$ its mobility, and $s_i(t) = \pm 1$ its polarity that flips sign at rate $\alpha/2$. In addition, $X$ denotes the position of the object that evolves according to
\begin{equation}\label{eq:X_motion}
    \dot{X} = \mu_{\mathrm{obj}}\sum^N_{i=1}V'(x_i-X),
\end{equation}
where $\mu_\mathrm{obj}$ is the object's mobility.

We note that our model assumes conservative interactions between active particles and the penetrable object, but the interactions may also be dominated by friction. Even in that case, we still find that the object exhibits symmetry-breaking motility, as detailed in Sec.~\ref{ssec:friction_case}.

\section{Negative drag} \label{sec:drag}

We first calculate the force applied by the RTPs on the object dragged at constant velocity $v$. For convenience, we adopt the frame of reference fixed to the object ($x_i \to x_i + X$). Then Eq.~\eqref{eq:RTP_motion} changes to
\begin{equation} \label{eq:RTP_motion_obj_frame}
    \dot{x}_i = -\mu\,V'(x_i) - v + u\,s_i(t)\quad\text{for $i \in \{1,\ldots,N\}$}.
\end{equation}
It is straightforward to convert this to the equations for the densities $\rho_\pm$ of the right- and left-moving RTPs:
\begin{align} \label{eq:rho+-}
	\partial_t \rho_+ = -\partial_x\{[F_\mathrm{eff}(x)+u]\rho_+\}+{\alpha\over2}(\rho_--\rho_+), \nonumber\\
    \partial_t \rho_- = -\partial_x\{[F_\mathrm{eff}(x)-u]\rho_-\}+{\alpha\over2}(\rho_+-\rho_-),
\end{align}
where $F_\mathrm{eff}\equiv-\mu\partial_x V-v$ is the effective force experienced by each RTP in the object frame. Using the total density $\rho \equiv \rho_+ + \rho_-$ and the polarization $\Delta \equiv \rho_+ - \rho_-$, Eq.~\eqref{eq:rho+-} can be rewritten as
\begin{align}
\partial_t \rho=-\partial_x J,\quad\quad
        \quad\;\;\;\quad J\equiv F_\mathrm{eff}\rho +u\Delta,
        \nonumber\\
        \partial_t \Delta=-\partial_x J_\Delta-\alpha \Delta,\;\;\;
        J_\Delta\equiv F_\mathrm{eff}\Delta +u\rho,
    \label{eq:rho&Delta}
\end{align}
where $J$ and $J_\Delta$ are the density and the polarization currents, respectively. Then, solving Eq.~\eqref{eq:rho&Delta} for the steady state ($\partial_t\rho = \partial_t\Delta = 0$), the drag force can be calculated as
\begin{align} \label{eq:Fobj}
    F_\mathrm{obj}(v) = \int_0^L dx\, \rho_\mathrm{s}(x;v)\,V'(x),
\end{align}
where $\rho_\mathrm{s}(x;v)$ denotes the steady-state density profile.

Remarkably, as illustrated in Fig.~\ref{fig:main}(b), the RTPs can pile up behind the object, in contrast to the case of a passive fluid where the particles always accumulate in front of the object. This implies that the force applied by the RTPs on the object is in the direction of motion: The RTPs exert a negative drag.

Let us denote by $\bar\rho \equiv N/L$ the mean density of the RTPs and by $l_p \equiv 2u/\alpha$ their persistence length. When $v$ is small, the drag force can be linearized as $F_\mathrm{obj} \simeq -(\bar\rho l_p/\mu) a_1 v$, where $a_1$ is the dimensionless drag coefficient. In the limit $L\to\infty$, the coefficient can be expressed as a function of the dimensionless parameters $\lambda \equiv \Lambda/l_p$ (rescaled object size) and $f\equiv \mu F/u$ (rescaled object-RTP repulsion):
\begin{align}
a_1(f,\lambda) = \frac{2}{f} \sinh\left(\frac{f\lambda}{1-f^2}\right)-\frac{\lambda \left(2-f^2+f^4\right)}{\left(1-f^2\right)^2}.
\end{align}
The boundary of the negative drag regime ($a_1 < 0$) is indicated by the dotted line in Fig.~\ref{fig:main}(c). This shows that $a_1 < 0$ requires both $\lambda$ and $f$ to be small enough.

\begin{figure*}
\includegraphics[width=\textwidth]{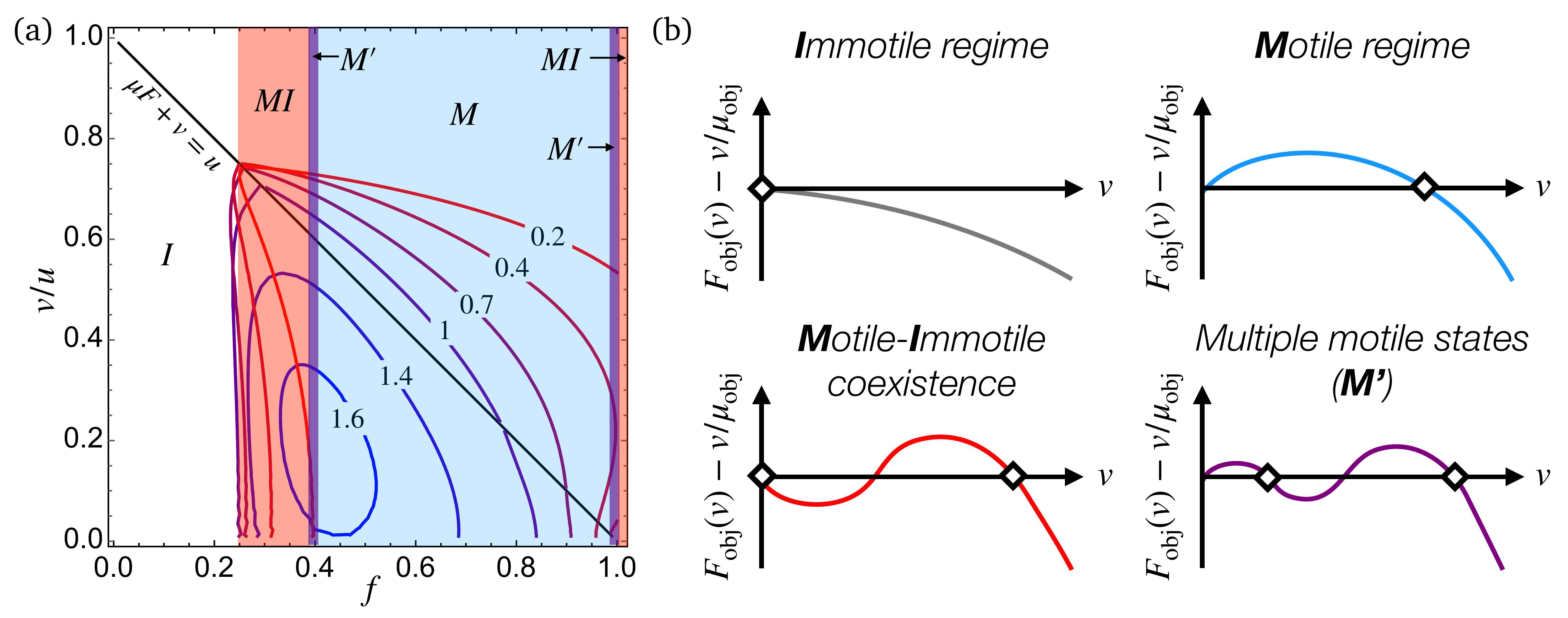}
\vspace{-0.25in}
\caption{\label{fig:MFT} (a) Self-consistent velocities of the object as the repulsion strength $f$ is varied for a fixed object size $\lambda$. The intervals of $f$ corresponding to each dynamical regime at $\lambda = 0.2$ are indicated by shaded areas. (b) Behavior of the net force on the object as a function of the object velocity $v$ for each dynamical regime.}
\end{figure*}

These properties can be understood intuitively as follows. The RTPs tend to move in the same direction even after penetrating into the object. Since they slow down inside the object, their density has to increase to keep the current uniform in the steady state. Thus, in contrast to the passive particles, whose density is always lower inside an object, the RTPs accumulate and form a high-density region at the object's surface, as illustrated in Fig.~\ref{fig:main}(b). When the object is static, the same number of particles accumulate on both sides. However, if the object moves, the magnitude $|F_\mathrm{eff}|$ of the effective repulsion is stronger behind the object than in front. This means that the RTP finds it easier to cross the object from the front to the rear than the other way around. Thus, the RTPs tend to accumulate more on the rear side of the object, inducing the negative drag. We note that this mechanism would work only when the object size is smaller than or comparable to the persistence length, so that the RTP keeps its direction of motion as it crosses the object. Moreover, the RTP-object repulsion should be weak enough to allow a sufficiently large flux between the two sides of the object. These are the reasons why the negative drag requires small $\lambda$ and $f$.

\section{Phase transitions} \label{sec:transitions}
\subsection{Mean-field predictions}

Thus far, we have assumed that the object moves at constant velocity. However, what would be the steady-state velocity of the object if allowed to move freely? Let us revisit Eq.~\eqref{eq:X_motion} describing the object motion. Assuming that the RTPs instantaneously relax to the steady state for a given object velocity $v = \dot{X}$, Eq.~\eqref{eq:X_motion} can be rewritten as a self-consistency equation
\begin{align} \label{eq:selfcon}
    F_{\mathrm{tot}}(v) \equiv F_{\mathrm{obj}}(v)-\frac{1}{\mu_{\mathrm{obj}}}v = 0.
\end{align}
Then the solutions of the above equation satisfying the stability condition $F'_\mathrm{obj}(v)<1/\mu_\mathrm{obj}$ approximate the steady-state object velocity. In this scheme, $v$ plays the role of the mean field for every RTP. Thus we may call Eq.~\eqref{eq:selfcon} the mean-field theory.

\begin{figure*}
\includegraphics[width=\textwidth]{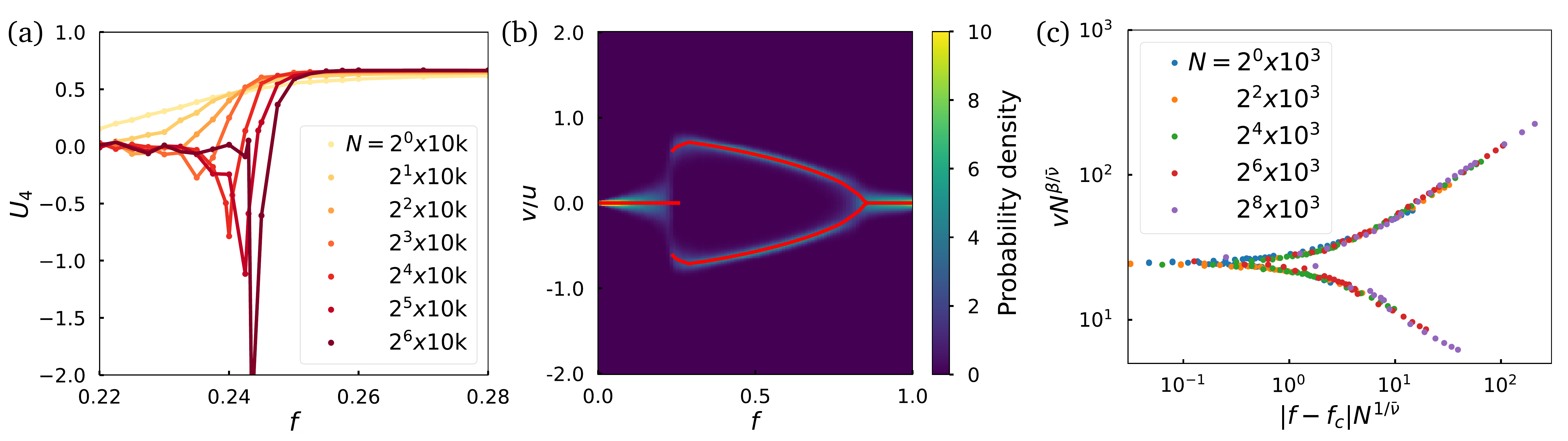}
\vspace{-0.15in}
\caption{\label{fig:transition} Analysis of the phase transitions of the object's dynamical state. (a) The Binder cumulant $U_4$ exhibits the hallmarks of a discontinuous transition. (b) Simulation confirms the bifurcations of the steady-state velocity as $f$ is varied. At $N = 3\times 10^4$, the results are in good agreement with the mean-field prediction (solid lines). (c) The continuous transition at $f_c \approx 0.81$ exhibits characteristics of the mean-field Ising universality class ($\beta=1/2$ and $\bar\nu = 2$). We use $\lambda = 1$ in all the results.}
\end{figure*}

Depending on the types of stable solutions, the steady-state object motion can be classified into four regimes, as shown in Fig.~\ref{fig:main}(c). (i) In the immotile (I) regime, $v = 0$ is the only stable solution. Here the object diffuses without any persistent traveling. (ii) In the motile (M) regime, $v > 0$ is the only stable solution. Here the object always travels persistently in a single direction. (iii) The motile-immotile (MI) coexistence regime has stable solutions at both $v = 0$ and $v > 0$. The object vacillates between the motile state and the immotile (diffusing) state. (iv) The regime of multiple motile states (M$^\prime$) has two stable positive solutions. The object vacillates between two different traveling velocities.~\footnote{For the range of parameters we checked ($0 \le f \le 1$ and $0 \le \lambda \le 2$), one of the nonzero solutions is very close to zero ($v/u$ is of order $10^{-2}$)} By the symmetry of the system, $-v$ is a stable solution of Eq.~\eqref{eq:selfcon} if $v$ is. 

For more details about how these regimes differ from each other, see Fig.~\ref{fig:MFT}. All solutions of Eq.~\eqref{eq:selfcon} for various $f$ and $\lambda$ are shown by contours in Fig.~\ref{fig:MFT}(a). The diagonal line $\mu F + v = u$ marks the boundary above which the RTPs approaching the object from behind cannot overtake the object. The line is important for determining which boundary conditions should be used in the mean-field theory, as detailed in Appendix~\ref{appdx:A}. Meanwhile, the behaviors of the left-hand side of Eq.~\eqref{eq:selfcon} as a function of $v$ are schematically illustrated for each dynamical regime in Fig.~\ref{fig:MFT}(b). The stable solutions are marked with diamonds.

This system is invariant under reflection about the object center, so it has the $Z_2$ symmetry. Any dynamical regime with a nonzero stable solution indicates that the symmetry is spontaneously broken. This implies the existence of phase transitions between the immotile and the motile regimes in the thermodynamic limit defined as $N \to \infty$ with $L$ and $N \mu_\mathrm{obj}$ fixed, which ensures that the right-hand side of Eq.~\eqref{eq:X_motion} converges to a finite value as $N \to \infty$.

Figure~\ref{fig:main}(c) indicates that there are two types of transitions between the immotile and the motile states. Along the thick black curve between the two white circles, the motile regime is in direct contact with the immotile regime. This curve marks a continuous transition. Indeed, in the vicinity of the thick black curve, the total force on the object can be expanded as
\begin{align} \label{eq:Ftot_exp}
    F_\mathrm{tot} \simeq \frac{\bar{\rho}l_p u}{\mu}\left\{-\left[a_1(f,\lambda,L)+\gamma\right]\frac{v}{u}+a_3(f,\lambda,L)\left(\frac{v}{u}\right)^3\right\},
\end{align}
where the even-order terms in $v$ do not appear because of the $Z_2$ symmetry, and $\gamma \equiv \mu/(\bar\rho\,l_p\,\mu_\mathrm{obj})$ is the rescaled friction coefficient of the object, which stays finite in the limit $N \to \infty$ because $\bar\rho \sim N$ and $\mu_\mathrm{obj} \sim 1/N$. We have fixed $\gamma = 0.1$ throughout this study, including Fig.~\ref{fig:main}(c). This amounts to fixing the mobilities $\mu$ and $\mu_\mathrm{obj}$ and the RTP properties $u$ and $\alpha$ while varying the object porosity $F$ and size $\Lambda$. Explicit calculations of $a_1$ and $a_3$ are given in Appendix~\ref{appdx:B}.

Along the thick black line shown in Fig.~\ref{fig:main}(c), the dimensionless coefficients of Eq.~\eqref{eq:Ftot_exp} are given by $a_1 = -\gamma$ and $a_3 < 0$. For a given value of $\lambda$, we denote the value of $f$ satisfying the condition $a_1 = -\gamma$ by $f_c(\lambda,L)$, at which a continuous transition occurs with the critical behavior $v \sim |f-f_c|^\beta$ with $\beta = 1/2$.

Meanwhile, we expect there to be a discontinuous transition line in each of the two MI regimes shown in Fig.~\ref{fig:main}(c). While both motile and immotile states are possible at finite $N$, we expect one of the two states to be exponentially more likely as $N$ grows. Also, there are two multicritical points located at the junctions between the critical line and the discontinuous transition lines, indicated by white circles in Fig.~\ref{fig:main}(c).

\subsection{Numerical results}

To verify the existence of discontinuous and continuous transitions, we ran extensive simulations of Eqs.~\eqref{eq:V_def}--\eqref{eq:X_motion} and examined the steady-state statistics of the system, with the results shown in Fig.~\ref{fig:transition} for $\lambda = 1$. As shown in Fig.~\ref{fig:main}(c), the mean-field theory predicts that varying $f$ along the $\lambda = 1$ line produces a discontinuous transition somewhere within the MI regime and a continuous transition at the critical line. 

In the heat map shown in Fig.~\ref{fig:transition}(b), the colors indicate the probability density of the rescaled object velocity $v/u$ for a given value of $f$. With the object mobility scaling as $\mu_\mathrm{obj} \sim 1/N$, one can expect the mean-field theory to become more accurate as $N$ grows because the dynamics of the object becomes slower compared to the relaxation of the RTPs. At $N=30\;000$ RTPs, the result already seems to be in good agreement with the mean-field predictions (solid curves).

The red curves clearly indicate the existence of a discontinuous transition. To verify this, in Fig.~\ref{fig:transition}(a) we plot the Binder cumulant~\cite{BinderZPB1981,*BinderPRL1981} $U_4 \equiv 1 - \langle v^4 \rangle/(3\langle v^2 \rangle^2)$ as a function of $f$ for various values of $N$. As $N$ increases, $U_4$ develops a dip which becomes narrower and deeper, a clear hallmark of a discontinuous transition.

In Fig.~\ref{fig:transition}(c) we present a finite-size scaling (FSS) analysis of the continuous transition behavior observed at $f_c \approx 0.81$. Using the FSS form
\begin{align}
v = N^{-\beta/\bar{\nu}}\,\Phi((f-f_c)N^{1/\bar\nu})
\end{align}
with the mean-field Ising critical exponents $\beta = 1/2$ and $\bar{\nu} = 2$, all the data obtained at different values of $N$ collapse onto a single curve. This implies that the critical phenomena are of the mean-field Ising universality class. We note that this is a natural consequence of the time-scale separation underlying the mean-field assumption.

Why do we observe such behavior, even though the system is 1D? This is because, via Eqs.~\eqref{eq:V_def}--\eqref{eq:X_motion} with the scaling $\mu_\mathrm{obj} \sim 1/N$, each RTP is coupled to the mean field of all the other RTPs. Via interactions with the object, all the RTPs are effectively coupled to each other, which resembles the all-to-all Ising model for which the mean-field theory is known to be exact.


\section{Effective diffusion}
\label{sec:diffusion}

\begin{figure}
\includegraphics[width=\columnwidth]{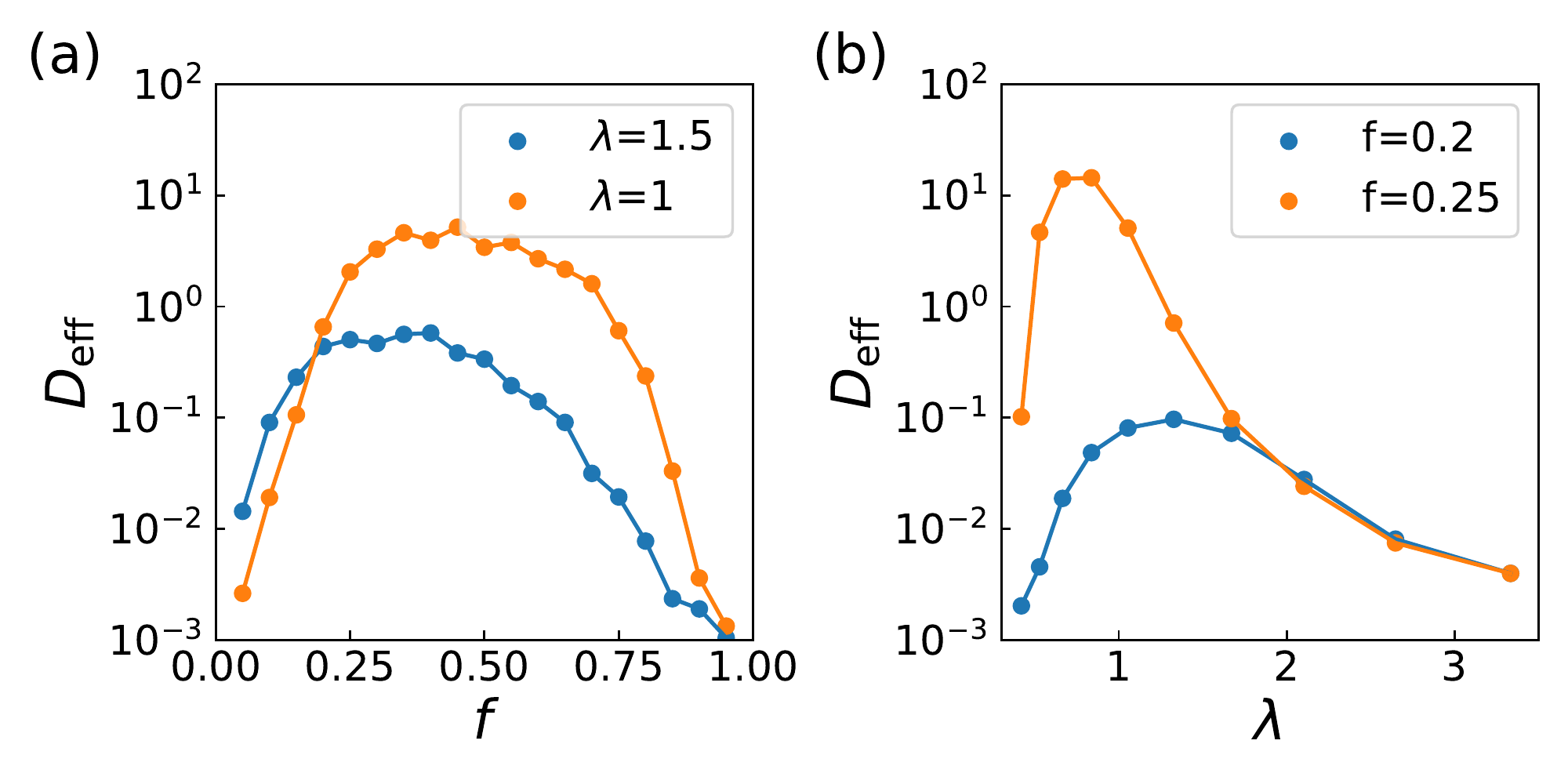}
\vspace{-0.3in}
\caption{\label{fig:Deff} Nonmonotonic behaviors of the effective diffusion coefficient of the object as (a) the object-RTP repulsion $f$ is varied and (b) the object size $\lambda$ is varied.}
\end{figure}

When the time-scale separation is not complete, the mean-field assumptions underlying our discussion so far are not strictly valid. Due to microscopic fluctuations of active particle density, the object dynamics becomes diffusive in the long-time limit. Then we can define the effective diffusion coefficient $D_\mathrm{eff} \equiv 
	\lim_{t\to\infty}\langle [X(t) - X(0)]^2\rangle/(2t)$. The dependence of $D_\mathrm{eff}$ on $f$ and $\lambda$ can be guessed from Fig.~\ref{fig:main}(c). Since $D_\mathrm{eff}$ is proportional to the product of the object velocity and its persistence length, we expect it to be the largest in the motile regime and the smallest in the immotile regime. Our numerics indeed confirm this intuition. In Fig.~\ref{fig:Deff}, we show the behaviors of $D_\mathrm{eff}$ as functions of $f$ and $\lambda$, respectively. These indicate that the diffusivity of a penetrable object immersed in an active fluid exhibits a nonmonotonic behavior as the porosity or the size of the object changes. The latter phenomenon is similar to \cite{PattesonSoftMatter2016}, but our mechanism is completely different.
 
\section{Extension to other systems} \label{sec:generalization}

Our findings thus far are limited to conservative RTP-object interaction in 1D active fluids. In this section we demonstrate that our results can be extended to more general systems.

\subsection{Perturbative solution for general potentials} \label{ssec:general_potl}

Here we show that, when the RTP-object interaction $V(x)$ is weak, the negative drag emerges not only for the triangular potential but also for general symmetric potentials. Towards this aim, we revisit Eq.~\eqref{eq:rho&Delta}, treating $V(x)$ as an arbitrary even function whose value is nonzero only for $-\Lambda/2 \le x \le \Lambda/2$. Then, assuming $V$ to be small, we solve the equation for the steady state perturbatively up to the linear order in the object velocity $v$. 

From here on, we drop the function argument if the meaning is clear. Let us begin with writing $\rho\approx \rho^{(0)}+v\rho^{(1)}$ and $J \approx vJ^{(1)}$, noting that $J = 0$ when the object is at rest ($v = 0$). In order $v^0$, the steady-state conditions are given by
\begin{align}
	\label{eq:zeroth1}
	0&=-\mu V'\rho^{(0)}+u\Delta^{(0)}
	\\
	\label{eq:zeroth2}
	0&=-\partial_x\left(-\mu V'\Delta^{(0)}+u\rho^{(0)} \right)-\alpha\Delta^{(0)}.
\end{align}
Eliminating $\Delta^{(0)}$, we have
\begin{align}
	0=\left(\mu^2 V'^2-u^2\right)\partial_x\rho^{(0)}+\left(2\mu^2 V'V''-\alpha\mu V'\right)\rho^{(0)}.
\end{align}
For small $V$, this equation is solved perturbatively as
\begin{align} \label{eq:rho0_soln}
	\rho^{(0)}\approx\rho_b\left(1-\frac{\mu\alpha}{u^2}V+\frac{\mu^2}{u^2}V'^2+\frac{\mu^2\alpha^2}{2u^4}V^2\right),
\end{align}
where $\rho_b$ is the bulk density of RTPs outside the object. To express $\rho_b$ in terms of the mean density $\bar\rho$, we use the normalization condition $\bar\rho L = \int_0^L dx\, \rho^{(0)}$, obtaining
\begin{align}
	\bar\rho L\approx\rho_b L+\rho_b\left(-\frac{\mu\alpha}{u^2}\overline{V}+\frac{\mu^2}{u^2}\overline{V'^2}+\frac{\mu^2\alpha^2}{2u^4}\overline{V^2}\right),
\end{align}
where $\overline{X} \equiv \int_{-\Lambda/2}^{\Lambda/2}dx\, X$. Since $L \gg 1$, this implies
\begin{align} \label{eq:rhob_soln}
	\rho_b\approx \bar\rho - \frac{\bar\rho}{L}\left(-\frac{\mu\alpha}{u^2}\overline{V}+\frac{\mu^2}{u^2}\overline{V'^2}+\frac{\mu^2\alpha^2}{2u^4}\overline{V^2}\right).
\end{align}
Using this in Eq.~\eqref{eq:rho0_soln} completes the solution for $\rho^{(0)}$.

Let us proceed to order $v$, where the steady-state conditions are given by
\begin{align}
	\label{eq:first1}
	J^{(1)}&=-\mu V'\rho^{(1)}-\rho^{(0)}+u\Delta^{(1)},
	\\
	\label{eq:first2}
	0&=-\partial_x\left(-\mu V'\Delta^{(1)}-\Delta^{(0)}+u\rho^{(1)}\right)-\alpha\Delta^{(1)}.
\end{align}
Eliminating $\Delta^{(0)}$ using Eq.~\eqref{eq:zeroth1} and $\Delta^{(1)}$ using Eq.~\eqref{eq:first1}, Eq.~\eqref{eq:first2} implies
\begin{align}
	\label{eq:first3}
	&\left(\mu^2V'^2-u^2\right)\partial_x\rho^{(1)}+\left(2\mu^2V'V''-\alpha\mu V'\right)\rho^{(1)}
	\nonumber\\
	&+\partial_x\left[\mu V'\left(J^{(1)}+2\rho^{(0)}\right)\right]-\alpha\left(J^{(1)}+\rho^{(0)}\right)=0.
\end{align}
To solve this perturbatively, we write $\rho^{(1)}\approx \rho^{(1)}_0+\rho^{(1)}_1$ and $J^{(1)}\approx J^{(1)}_0+J^{(1)}_1$, where $\rho^{(1)}_n$ and $J^{(1)}_n$ denote the solution at order $V^n$. At order $V^0$, Eqs.~\eqref{eq:rho0_soln}, \eqref{eq:rhob_soln}, and \eqref{eq:first3} imply
\begin{align}
	u^2\partial_x\rho^{(1)}_0+\alpha\left(J^{(1)}_0+\bar\rho\right)=0,
\end{align}
which is solved by
\begin{align}
	\rho^{(1)}_0 =-\frac{\alpha}{u^2}\left(J^{(1)}_0+\bar\rho\right)x+\rho^{(1)}_0(-\Lambda/2).
\end{align}
Due to the periodic boundary condition, the coefficient of $x$ must vanish, yielding $J^{(1)}_0=-\bar\rho$. Then, the normalization condition $\int_0^L dx\,\rho^{(1)} = 0$ implies $\rho^{(1)}_0=0$.

At order $V$, Eqs.~\eqref{eq:rho0_soln}, \eqref{eq:rhob_soln}, and \eqref{eq:first3} imply
\begin{align}
	u^2\partial_x\rho^{(1)}_1=\mu V''\bar\rho-\alpha J^{(1)}_1-\alpha\left(-\frac{\bar\rho\mu\alpha}{u^2}+\frac{\bar\rho\mu\alpha}{u^2L}\overline{V} \right),
\end{align}
whose solution is 
\begin{align}
	\rho^{(1)}_1&=\rho^{(1)}_1(-\Lambda/2)+\frac{\mu V'}{u^2}\bar\rho-\frac{\alpha}{u}J^{(1)}_1x\nonumber\\
	&\qquad+\frac{\bar\rho\mu\alpha^2}{u^4}\left(\mathcal{V}-\frac{\overline{V}}{L}x  \right),
\end{align}
where $\mathcal{V}(x)\equiv\int_{-\Lambda/2}^xdx'\,V(x')$. The periodic boundary condition implies $J^{(1)}_1=0$, and the density normalization $\int_0^L dx\,\rho^{(1)} = 0$ yields
\begin{align}
	\rho^{(1)}_1(-\Lambda/2)=\frac{\bar\rho\mu\alpha^2}{u^4}\left(\frac{1}{2}\overline{V}-\frac{1}{L}\int^L_0 dx\,\mathcal{V} \right).
\end{align}
Thus, we obtain
\begin{align} \label{eq:rho1_soln}
	\rho^{(1)}_1&=\frac{\bar\rho\mu\alpha^2}{u^4}\left(\frac{1}{2}\overline{V}-\frac{1}{L}\int^L_0 dx\,\mathcal{V} \right)+\frac{\mu V'}{u^2}\bar\rho\nonumber\\
	&\qquad+\frac{\bar\rho\mu\alpha^2}{u^4}\left(\mathcal{V}-\frac{\overline{V}}{L}x  \right).
\end{align}

Finally, using the above results, the drag exerted on the object is obtained as
\begin{align}
	F_{\mathrm{obj}}&=\int^L_0 dx\,\rho V'\approx v\int^L_0 dx\,\rho^{(1)}V'\approx v\int^L_0 dx\,\rho^{(1)}_1V' \nonumber\\
	&\approx v\int^L_0 dx\, \left(\frac{\bar\rho\mu}{u^2}V'^2+\frac{\bar\rho\mu\alpha^2}{u^4}V'\mathcal{V}-\frac{\bar\rho\mu\alpha^2\overline{V}}{u^4L}V'x \right)
	\nonumber\\
	&\approx v\left(\frac{\bar\rho\mu}{u^2}\overline{V'^2}- \frac{\bar\rho\mu\alpha^2}{u^4}\overline{V^2}\right),
\end{align}
where the last expression is derived via integration by parts and taking $L \to \infty$. Noting that the effective diffusion coefficient of the free RTP is given by $D_\mathrm{eff} = u^2/\alpha$, the Einstein relation yields the effective temperature of the RTPs $T_\mathrm{eff} = D_\mathrm{eff}/\mu = u^2/(\mu \alpha)$. Using this quantity as well as the persistence length $l_p \equiv 2u/\alpha$, the drag coefficient can be written as
\begin{align}
\gamma_\mathrm{drag} \equiv -\lim_{v\to 0}\frac{F_{\mathrm{obj}}}{v}\approx\frac{\bar\rho}{\mu}\frac{\overline{V^2}}{T_\mathrm{eff}^2} \left(1-\frac{l_p^2}{4}\frac{\overline{V'^2}}{\overline{V^2}}\right).
\end{align}
This result is very instructive about how the negative drag emerges in active fluids. In the equilibrium limit, which corresponds to $l_p \to 0$ with finite $T_\mathrm{eff}$, $\gamma_\mathrm{drag}$ is bound to be positive. However, since the activity of the RTPs produces finite $l_p$, $\gamma_\mathrm{drag}$ is reduced, becoming even negative when $l_p$ is sufficiently larger than the object length scale given by $\sqrt{\overline{V^2}/\overline{V'^2}}$. We have thus shown that the negative drag can emerge for generic RTP-object interaction potentials.

\subsection{Frictional RTP-object interactions} \label{ssec:friction_case}

We focused on the case where the RTP and the object interact via conservative forces. In this section, we demonstrate that negative drag can also emerge from nonconservative interactions. Towards this goal, we examine the case where an RTP crossing the object experiences a constant kinetic friction in the direction opposite to its relative velocity. In the overdamped limit, the equation of motion for such RTPs reads
\begin{align}
	\dot{x}_i =-\mu F(x_i-X)\,s_i(t)+u\,s_i(t) \;\; \text{for}\; i\in \{1,...,N\},
\end{align}
where $F(x)$ is the friction force inside the object defined as
\begin{align}
	F(x)=\begin{cases}
		F \;\; &\text{if} \; x\in [-\Lambda/2,\Lambda/2]\\
		0 \;\; &\text{otherwise}.
	\end{cases}
\end{align}
In the reference frame of the moving object ($x_i\to x_i+X$), the equation becomes
\begin{align}
	\dot{x}_i=-v+[u-f(x)]s_i(t) \;\; \text{for}\; i\in \{1,...,N\},
	\label{seq:fric_motion}
\end{align}
where $f(x)\equiv \mu F(x)$. We also introduce the notation $f \equiv \mu F$. 

 Now it is straightforward to convert Eq.~\eqref{seq:fric_motion} to the master equation
\begin{align}
		\partial_t\rho_+=-\partial_x[(-v+u-f(x))\rho_+]+\frac{\alpha}{2}(\rho_--\rho_+),\nonumber\\
		\partial_t\rho_-=-\partial_x[(-v-u+f(x))\rho_-]+\frac{\alpha}{2}(\rho_+-\rho_-).
\end{align}
In terms of the total density $\rho \equiv \rho_+ + \rho_-$ and the polarization $\Delta = \rho_+ - \rho_-$, the equations can be rewritten as
\begin{align}
		\partial_t\rho&=-\partial_x J, &\text{$J\equiv -v\rho+[u-f(x)]\Delta$},\nonumber\\
		\partial_t\Delta &=-\partial_x J_\Delta -\alpha \Delta, & \text{$J_\Delta\equiv[u-f(x)]\rho -v\Delta$}. 
\label{seq:master_eq_friction}
\end{align}

The force $F_{\text{obj}}$ exerted on the object by the RTPs is then calculated as
\begin{align}
	F_{\text{obj}}=\int^{\Lambda/2}_{-\Lambda/2} dx\, F(x)\,[\rho_+(x)-\rho_-(x)]=\int^{L}_{0} dx\, F(x)\,\Delta(x).
	\label{seq:fric_obj_force}
\end{align}

As detailed in Appendix~\ref{appdx:C}, the resulting form of exerted force in the limit $L\to\infty$ is
\begin{align}\label{eq:fricFobj}
	F_{\text{obj}}=\frac{\bar{\rho}f^2}{\mu\alpha}\left\{
	1-\exp\left[
	-\frac{\alpha v\Lambda}{(u-f)^2-v^2}
	\right]
	\right\}.
\end{align}
Therefore, for $v<u-\mu F$, the object always experiences negative drag.

\begin{figure}
\includegraphics[width=0.35\textwidth]{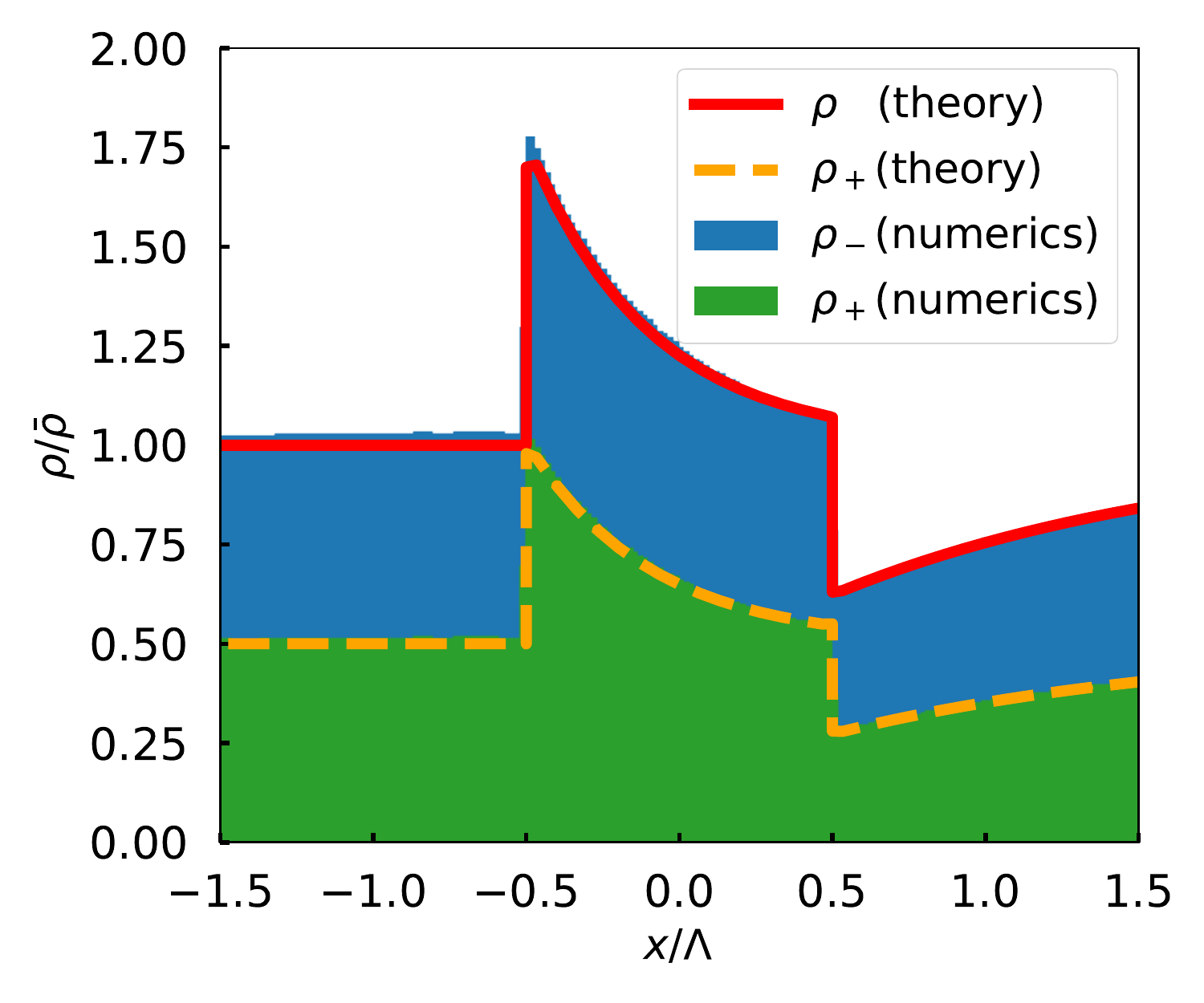}
\caption{\label{fig:fig5} Density profile of the RTPs around a symmetric object moving to the right at velocity $v/u = 0.2$ which interacts with RTPs through the constant kinetic friction. The numerics refer to the results of a particle-based simulation using $50\;000$ RTPs. The portions accounted for by the left-moving ($\rho_-$) and the right-moving ($\rho_+$) RTPs are distinguished using different shades. The theoretical predictions (solid lines) are in good agreement with the numerical results (shaded areas).}
\end{figure}

To validate our theory, we conducted extensive numerical simulations of Eq.~\eqref{seq:fric_motion}. Indeed, the observed density profiles of the RTPs shown in Fig.~\ref{fig:fig5} are in good agreement with our theoretical predictions.

Remarkably, when the RTP-object interaction is purely based on the constant kinetic friction, negative drag occurs irrespective of the size of the object. This is in stark contrast to the case of conservative RTP-object interaction, where the object size had to be smaller or comparable to the persistence length of the RTPs for negative drag to occur.

Why are the two cases so different? When the interaction is conservative, $F_\mathrm{obj} > 0$ is satisfied when there are more particles behind the object than in front of the object; in other words, the position distribution of the RTPs should be biased in the right way, which requires that most RTPs completely cross the object in a short time, so that they can see the difference between the two sides of the object. In contrast, when the RTP and the object interact via kinetic friction, $F_\mathrm{obj} > 0$ only requires that there are more right-movers ($\rho_+$) than left-movers ($\rho_-$) inside the object [see Eq.~\eqref{seq:fric_obj_force}]. This mechanism is completely indifferent to how the two sides of the object differ from each other, so there is no need for the RTPs to cross the object in a short time. Hence, negative drag occurs regardless of the size of the object in this case.

\subsection{Extension to two-dimensional systems}
\label{ssec:two-dim}

While we focused on the model limited to the 1D case, here we demonstrate that the symmetry-breaking mechanism is also possible in 2D systems.

We consider an overdamped, penetrable disk of radius $R$ immersed in an active ideal gas on a 2D torus of size $L\times L$. The gas consists of $N$ active Brownian particles (ABPs), which do not interact with each other but only with the object via the conic potential
\begin{align}
	V(\bold{r}) = \begin{cases}
 	F(R-|\bold{r}|)	& \text{for $0 \le |\bold{r}|< R$,} \\
 	0			& \text{otherwise.}
 \end{cases}
\end{align}
We again assume that thermal noise is negligible compared to the other forces. With these assumptions, the equations of motion for each ABP are given by
\begin{align}
	&\dot{\bold{r}}_i=-\mu\nabla_{\bold{r}_i} V(\bold{r}_i-\bold{X})-u\hat{\bold{n}}_i(t)\;\;\; \text{for}\;i\in\{1,...,N\},\\
	&\dot{\theta}_i=\sqrt{2D_r}\eta_i(t),
\end{align}
where $\bold{r}_i$ is the position of the $i$-th ABP, $u$ its speed, $\mu$ its mobility, $\hat{\bold{n}}\equiv(\cos\theta_i,\sin\theta_i)$ its orientation, $D_r$ its rotational diffusion coefficient, and $\eta_i(t)$ the Gaussian white noise satisfying $\langle \eta_i(t) \rangle=0$ and $\langle \eta_i(t)\eta_j(t') \rangle=\delta_{ij}\delta(t-t')$. In addition, $\bold{X}$ denotes the position of the object and evolves according to
\begin{align}
	\dot{\bold{X}}=-\mu_{\text{obj}}\sum^N_{i=1}\nabla_{\bold{X}}V(\bold{r}_i-\bold{X}),
\end{align}
where $\mu_{\text{obj}}$ is the mobility of the object. 

For the simulation, we used the following parameters: $N=50\;000$, $u=5$, $D_r=\pi^2/3$, $\mu=5$, $\mu_{\text{obj}}=0.01$, $R=3$, and $L=15$.

\begin{figure}
\includegraphics[width=\columnwidth]{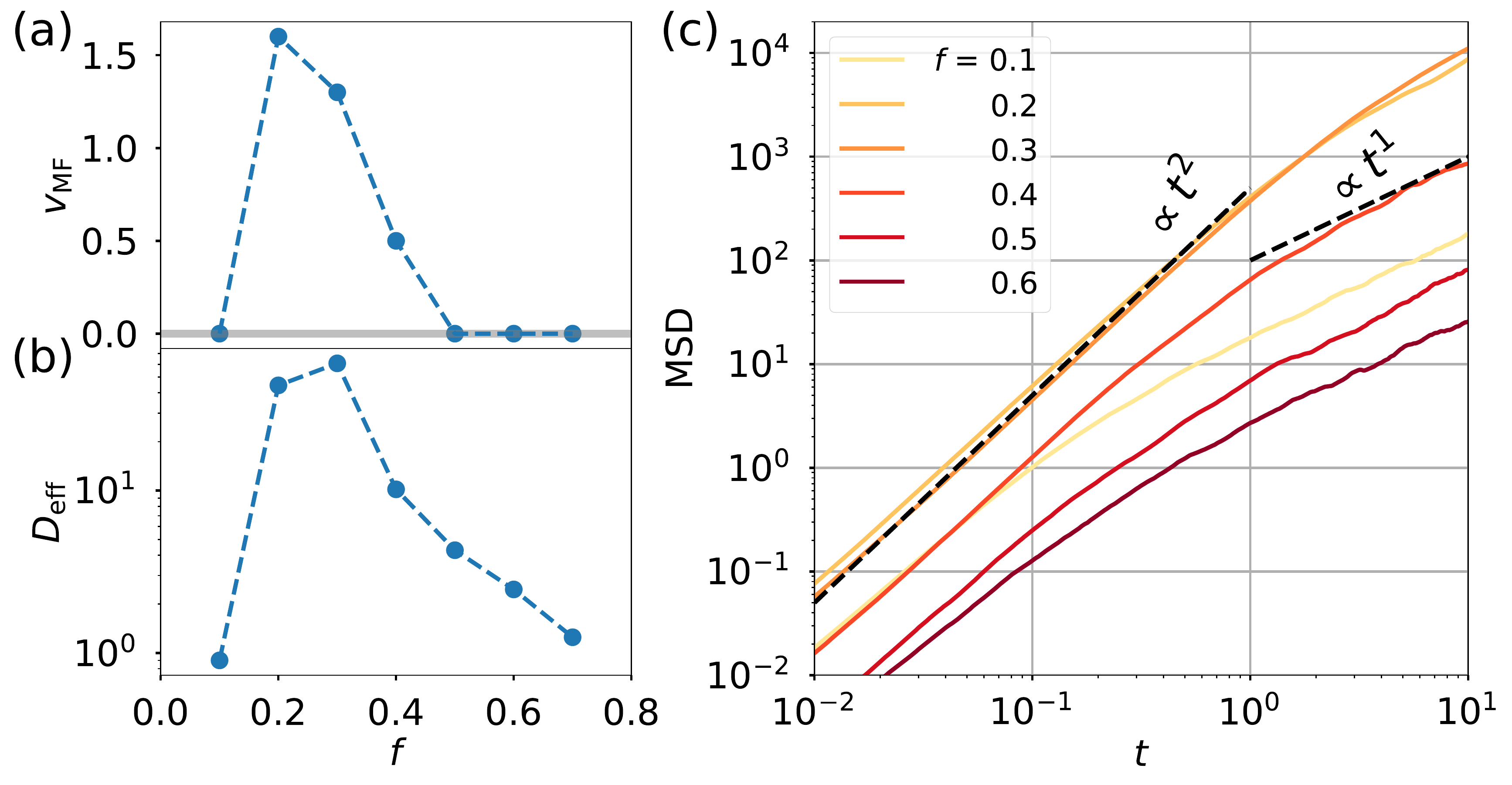}
\caption{\label{fig:fig6} (a) Mean-field predictions for the steady-state velocity of the object and (b) effective diffusion coefficient of the object as the repulsion strength $f$ is varied. The dashed lines are to guide the eye. (c) Mean square displacement of the object as the repulsion strength $f$ is varied, obtained by averaging over $200$ samples.}
\end{figure}

We first calculate the force applied by ABPs on the object when it is dragged at constant velocity $v\hat{\bold{e}}_x$, where $\hat{\bold{e}}_x$ is the unit vector in the $x$ direction. We adopt the frame of reference fixed to the object ($\bold{r}_i\to \bold{r}_i+\bold{X}$) for convenience. In the steady state, the drag force on the object is obtained as
\begin{align}
	\bold{F}_{\text{obj}}(v)=\int^R_0\int^{2\pi}_0 r\,dr\, d\theta\, \rho_s(\bold{r};v)\nabla_{\bold{r}}V(\bold{r}),
\end{align}
where $\rho_s(\bold{r};v)$ denotes steady-state density profile of the ABPs. Then we calculate the mean-field object velocity $v_{\text{MF}}$ using the self-consistency equation
\begin{align}
	\bold{F}_{\text{tot}}(v_{\text{MF}})\cdot \hat{\bold{e}}_x\equiv 
	\bold{F}_{\text{obj}}(v_{\text{MF}})\cdot \hat{\bold{e}}_x-\frac{1}{\mu_{\text{obj}}}v_{\text{MF}}=0.
\end{align}
As shown in Fig.~\ref{fig:fig6}(a), the mean-field assumption predicts that the object becomes motile ($v_\mathrm{MF} \neq 0$) only for an intermediate range of the repulsion strength $f$ in a manner similar to the 1D case.

\begin{figure}
\includegraphics[width=\columnwidth]{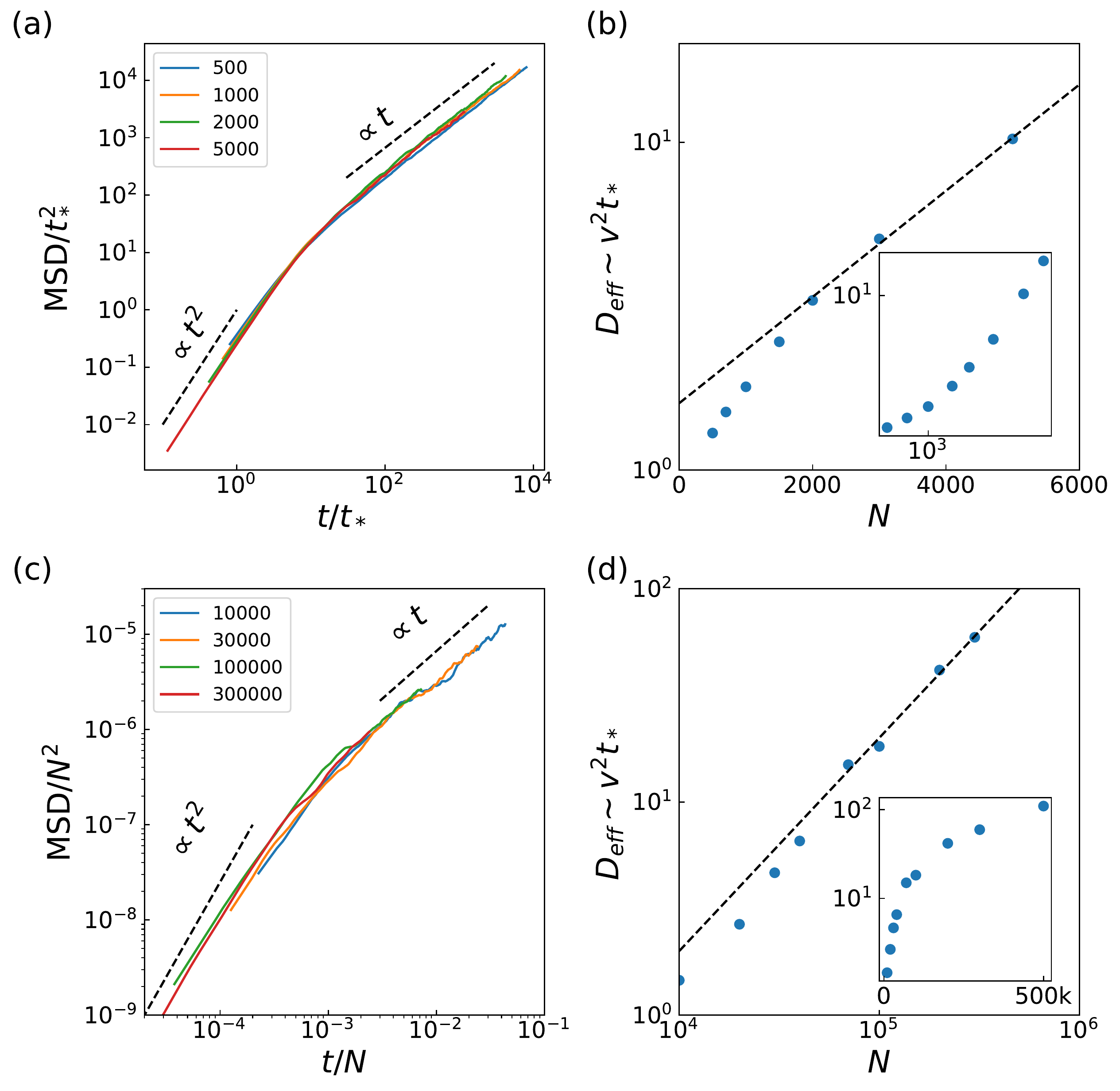}
\caption{\label{fig:fig7} (a) Mean square displacement of the passive object in the 1D case. The data collapse is done with the persistence time scale $t_* \sim \exp(0.0004N)$, which is also confirmed by (b) the scaling behavior of $D_\mathrm{eff}$ with respect to the number $N$ of RTPs. The inset shows the log-log version of the main semi-logarithmic plot. The corresponding results for the 2D case are shown in (c) and (d), with the inset being the semi-logarithmic version of the main log-log plot.}
\end{figure}

To check that the mean-field assumption correctly predicts the phenomenology, we let the object move freely and measure the effective diffusion coefficient defined as
\begin{align}
	D_{\text{eff}}\equiv \lim_{t\to\infty}\frac{\langle[\bold{X}(t)-\bold{X}(0)]^2 \rangle}{2t},
\end{align}
with the results shown in Fig.~\ref{fig:fig6}(b). We observe that $D_\mathrm{eff}$ exhibits a nonmonotonic dependence on $f$, which is consistent with the mean-field predictions of $v_\mathrm{MF}$.

Finally, we also check the mean square displacement (MSD) of the object
\begin{align}
\text{MSD}(t)\equiv \left\langle[\bold{X}(t+\tau)-\bold{X}(\tau)]^2\right\rangle,
\end{align}
where $\tau$ is some initialization time set to be longer than the relaxation time scale. The results, shown in Fig.~\ref{fig:fig6}(c), confirm that the enhanced diffusion coefficient in the intermediate range of $f$ is indeed due to the initial ballistic motion of the object, which must be the consequence of the negative drag.

Our results clearly show that the symmetry-breaking motility of passive penetrable objects in an active fluid is not limited to the 1D case but can generally occur in higher-dimensional systems. However, we also observe a notable difference between how the motion of the object crosses over from the ballistic regime to the diffusive regime in the 1D and the 2D cases. In the 1D case, the object can switch its direction only when many of the RTPs accumulated behind it move to the other side of the object. This is analogous to the magnetization of a ferromagnetic Ising system crossing the energy barrier and changing its sign. In the 2D case, the object can change its direction of motion via angular diffusion without such barrier-crossing events.

To demonstrate this difference, in Fig.~\ref{fig:fig7} we show the MSD of the object in the 1D and the 2D cases, varying the number $N$ of the RTPs while rescaling the object mobility according to $\mu_\mathrm{obj} \sim N^{-1}$. As $N$ is increased, the variance of the force exerted by the RTPs on the object scale as $N^{-1}$. For the 1D case, by analogy with the Arrhenius equation, the persistence time of the object scales as $\tau_* \sim \exp(cN)$. On the other hand, for the 2D case, the persistence time scales as $\tau_* \sim N$. These are indeed confirmed by the data collapses of the MSD in Figs.~\ref{fig:fig7}(a) and \ref{fig:fig7}(c) and the plots of the effective diffusion coefficient $D_\mathrm{eff}$ in Figs.~\ref{fig:fig7}(b) and \ref{fig:fig7}(d), which scales as $D_\mathrm{eff} \sim v^2t_*$. Note that changing $N$ does not affect the average speed $v$ of the object, so the $D_\mathrm{eff}$ and $t_*$ have the same scaling relationship with $N$.

\section{Summary and outlook} \label{sec:summary}
 
We described theoretically the steady-state dynamics of a 1D symmetric penetrable object immersed in an ideal gas of RTPs. We found that the drag coefficient of the object becomes negative when the object size and the object-RTP repulsion are both sufficiently small. In that case, the object moves persistently in a single direction by breaking the symmetry. Provided the complete time scale separation between the object and the RTPs, the steady-state velocity of the object exhibits discontinuous and continuous phase transitions, with the former involving the coexistence of multiple dynamical states and the latter exhibiting the mean-field Ising critical phenomena. Even if the time scale separation is not complete, these transitions increase the diffusion coefficient of the object by several orders of magnitude, hinting at the interesting possibility that properties of an active fluid can be dramatically altered by tuning the size of passive impurities.

Also we demonstrated that a 2D penetrable object also exhibits symmetry-breaking motility. Still, it remains to be checked whether the same mechanism is at work even for higher-dimensional objects with a hard core.

Finally, it would be interesting to explore applications to active engines~\cite{KrishnamurthyNP2016,MartinEPL2018,PietzonkaPRX2019,EkehPRE2020,LeePRE2020,FodorEPL2021} and collective phenomena involving multiple symmetric objects arising from the long-range interactions mediated by active particle currents~\cite{BaekPRL2018,GranekJSM2020}.

\begin{acknowledgments}
This work was supported by the National Research Foundation of Korea Grant funded by the Korean Government (Grant No. NRF-2020R1C1C1014436). Y.B. thanks Yariv Kafri, Alexander Solon, Nikolai Nikola, Xinpeng Xu, and Patrick Pietzonka for helpful comments.
\end{acknowledgments}

\appendix

\section{Derivation of the exerted force on the object}\label{appdx:A}

\begin{figure}[b]
\includegraphics[width=0.7\columnwidth]{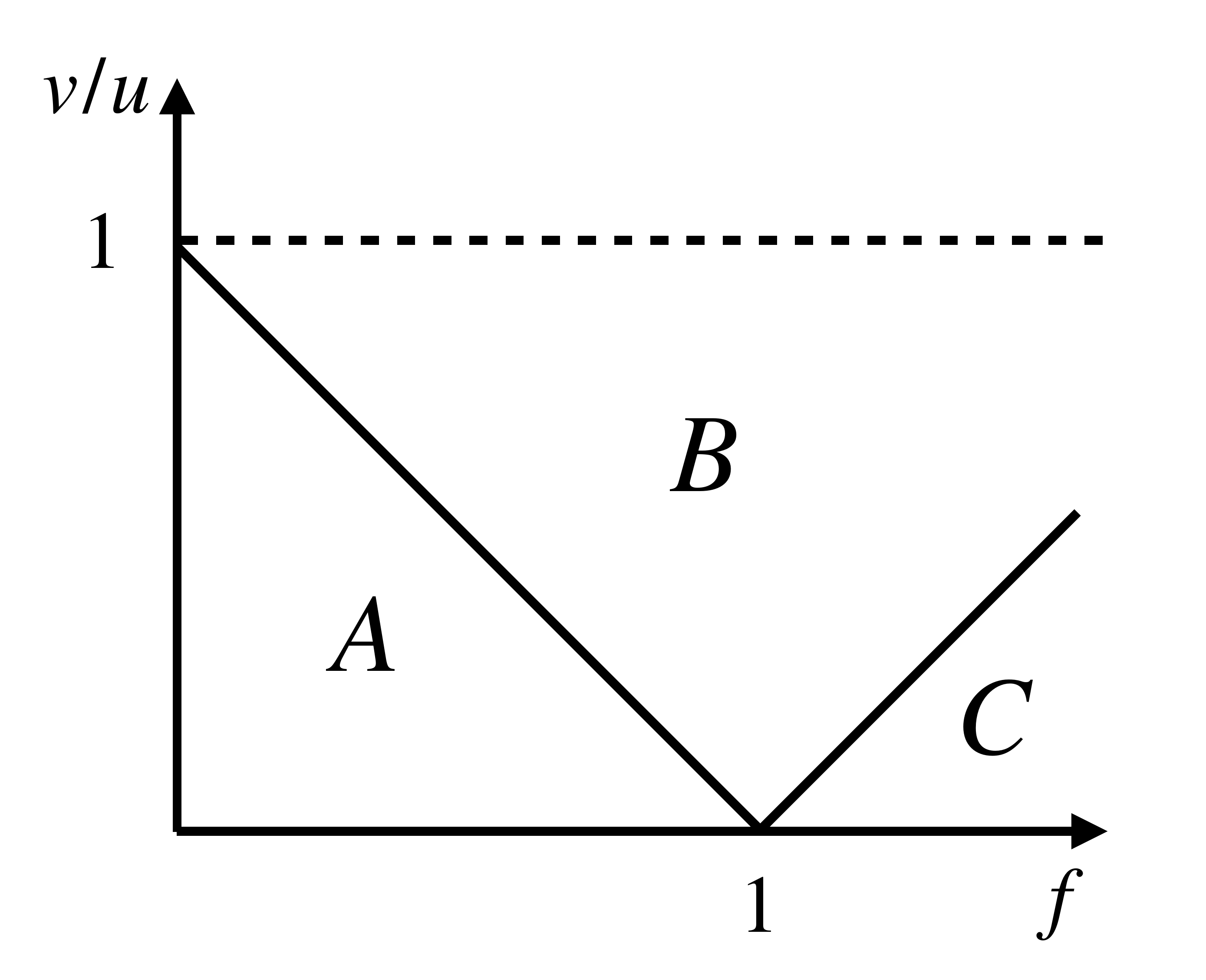}
\caption{\label{fig:fig8} Three regimes classified according to the penetrability of the object}
\end{figure}

We derive the expression for $F_\mathrm{obj}$ using the steady-state density of RTPs around the object moving at constant velocity $v$. Towards this goal, as pointed out in \cite{LeDoussalEPL2020}, we need to separately address three different cases described below and illustrated in Fig.~\ref{fig:fig8} (assuming that the object moves to the right with $v > 0$).
 
{\em Case A.} For $\mu F+v<u$, the RTPs can penetrate the object from both sides.
{\em Case B.} For $\mu F+v>u$ and $\mu F -v<u$, the right-moving RTPs cannot pass through the object whereas the left-moving RTPs can.
{\em Case C.} For $\mu F>u+v$, no RTPs can penetrate the object.

We neglect the case where the object is faster than the RTPs since such situations would not arise naturally from the object-RTP interactions. Moreover, we are not interested in case~C where the RTPs are bound to accumulate more in front of the object than behind, making the immotile state the only stable solution. Thus, we focus on cases~A and B.

\subsection{Force--current relationship}


Utilizing Eqs.~\eqref{eq:rho&Delta} and \eqref{eq:Fobj},
\begin{align}
F_\mathrm{obj}&=-\frac{1}{\mu} \int_0^L dx\, (J+v\rho-u\Delta)\nonumber\\
			&=-\frac{1}{\mu}\left(JL+vN-u\int_0^L dx\,\Delta\right).
\end{align}
In the steady state, Eq.~\eqref{eq:rho&Delta} implies
\begin{align}
    F_\mathrm{obj}&=-\frac{1}{\mu}\left[
    JL+vN-\frac{u}{\alpha}\int_0^L dx\, (\partial_x J_{\Delta})\right].
\end{align}
Since the system is periodic, the last term on the right-hand side is zero. Thus, we obtain
\begin{align}
    F_{\text{obj}}=-\frac{1}{\mu}(JL+vN),
    \label{B5}
\end{align}
which describes the force--current relationship in the object frame. We note that a similar expression was derived in \cite{NikolaPRL2016} for the laboratory frame of reference.

\subsection{Steady-state RTP density}\label{subsec:RTP_formula}

In order to obtain $J$, we first derive expressions for the steady-state density of the RTPs. In the steady state, the elimination of $\Delta$ in Eq.~\eqref{eq:rho&Delta} yields
\begin{align}
    \partial_x\{[u^2-F_\mathrm{eff}(x)^2]\,\rho(x)\}&-\alpha F_\mathrm{eff}(x)\,\rho(x)\nonumber\\
	&+ [\alpha+F_\mathrm{eff}'(x)]J=0.
\end{align}
Defining
\begin{align}
g(x) &\equiv [u^2-F_\mathrm{eff}(x)^2]\rho(x), \label{seq:g_def} \\
a(x) &\equiv \frac{\alpha F_\mathrm{eff}(x)}{u^2-F_\mathrm{eff}(x)^2}, \label{seq:a_def} \\
b(x) &\equiv [\alpha+F_\mathrm{eff}'(x)]J, \label{seq:b_def}
\end{align}
the equation can be rewritten as
\begin{align}
    g'(x)-a(x)\,g(x)+b(x)=0.
    \label{seq:g}
\end{align}
Since this is a first-order ordinary differential equation, its general solution is straightforwardly obtained as
\begin{align}
    g(x)&-g(c)\exp{\left\{ \int^x_c{dx_2\,a(x_2)} \right\}}\nonumber\\
    &=-\int^x_c{dx_1\,b(x_1)\exp{\left\{ -\int^{x_1}_{x}{dx_2\, a(x_2)} \right\}}},
    \label{seq:g_solution}
\end{align}
where $c$ is an arbitrary constant. Then, using the definition of $g(x)$, we can write
\begin{align}
    \rho(x)&=\frac{u^2-F_{\mathrm{eff}}(c)^2}{u^2-F_{\mathrm{eff}}(x)^2}\,\rho(c)\exp{\left\{ \int^x_c{dx_2\,a(x_2)} \right\}}\nonumber\\
    &~~-\frac{1}{u^2-F_{\mathrm{eff}}(x)^2}\int^x_c{dx_1\,b(x_1)\exp{\left\{ \int^x_{x_1}dx_2\,a(x_2) \right\}}}.
    \label{seq:density_formula}
\end{align}
From this expression, we learn the following.

\begin{figure}
\includegraphics[width=\columnwidth]{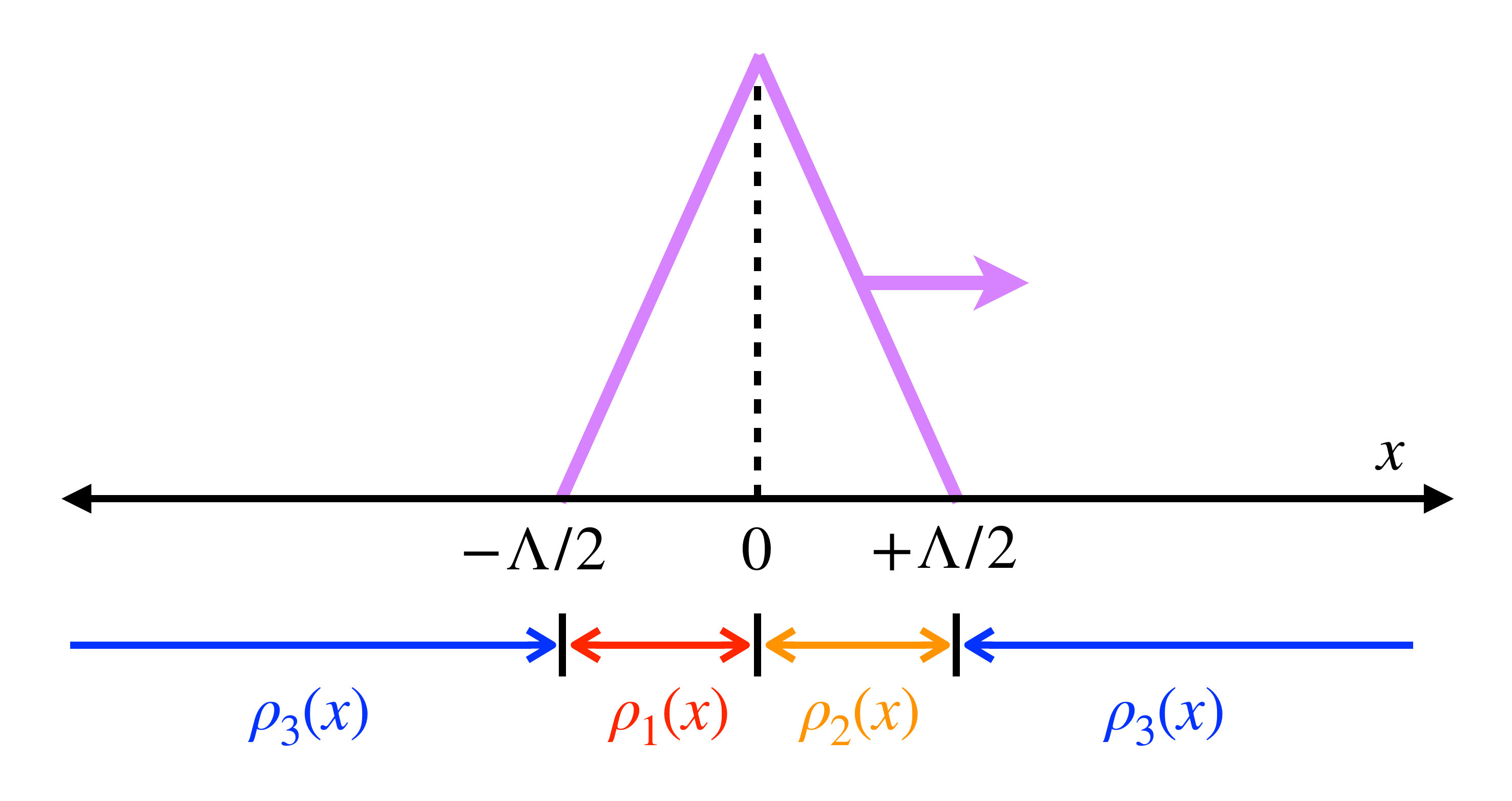}
\caption{\label{fig:fig9} RTP densities for respective region.}
\end{figure}

(i) Since $F_\mathrm{eff}(x)$ changes discontinuously at $x = \pm \Lambda/2$ and $x=0$, $b(x)$ has delta peaks at these locations. This implies that $\rho(x)$ has discontinuous jumps at 	the same locations. For this reason, as shown in Fig.~\ref{fig:fig9}, we set
\begin{align}
	\rho(x) = \begin{cases}
 	\rho_1(x) &\text{for $-\Lambda/2 < x < 0$,}\\
 	\rho_2(x) &\text{for $0 < x < \Lambda/2$,}\\
 	\rho_3(x) &\text{otherwise,}
 \end{cases}
\end{align}
and apply Eq.~\eqref{seq:density_formula} separately to $\rho_1$, $\rho_2$, and $\rho_3$ to calculate these functions.

(ii) The discontinuities of $F_\mathrm{eff}(x)$ should be understood as the limiting behaviors of some smoothly behaved effective force profile. More specifically, while we have assumed $F_\mathrm{eff}(x)$ to take only one of the three values $\pm \mu F-v$ and $-v$, the function actually takes all values in between, rapidly changing in the infinitesimal neighborhoods of $x = \pm \Lambda/2$ and $x=0$. In case~A, where $u > |F_\mathrm{eff}|$, such a continuous change of $F_\mathrm{eff}(x)$ never satisfies $u^2 - F_\mathrm{eff}^2(x) = 0$; thus, $\rho(x)$ as described by Eq.~\eqref{seq:density_formula} stays finite. In contrast, in case~B, $u^2 - F_\mathrm{eff}^2(x) = 0$ is achieved in the infinitesimal neighborhoods of $x = -\Lambda/2$ and $x = 0$; thus, Eq.~\eqref{seq:density_formula} implies that $\rho(x)$ may diverge to infinity at these locations. Hence extra care must be taken when dealing with the boundary conditions there.

\subsection{Solution for case A}

Let $x_0^+$ ($x_0^-$) indicate a point infinitesimally close to $x_0$ with $x_0^- < x_0 < x_0^+$. Then, using $c = -\Lambda/2^+$, $0^+$, and $\Lambda/2^+$ in Eq.~\eqref{seq:density_formula}, we obtain
\begin{align}
&\rho_1(x)
    =\rho(-\Lambda/2^+)\exp{\left[ -\frac{\alpha(\mu F+v)}{u^2-(\mu F+v)^2}\left(x+\frac{\Lambda}{2}\right) \right]} \nonumber\\
    &~~-\frac{J}{\mu F+v}\left\{ 1-\exp{\left[ -\frac{\alpha(\mu F+v)}{u^2-(\mu F+v)^2} \left( x+\frac{\Lambda}{2} \right)\right]} \right\},
    \label{B13}\\
&\rho_2(x)=\rho(0^+)\exp{\left[ \frac{\alpha(\mu F-v)}{u^2-(\mu F-v)^2} x\right]}\nonumber\\
	&\qquad\qquad+\frac{J}{\mu F-v}\left\{ 1-\exp{\left[ \frac{\alpha(\mu F-v)}{u^2-(\mu F-v)^2}x
    \right]} \right\},
    \label{B14}\\
&\rho_3(x)=\rho(+\Lambda/2^+)\exp{\left[ -\frac{\alpha v}{u^2-v^2}\left(x-\frac{\Lambda}{2}\right) \right]}\nonumber\\
&\qquad\qquad-\frac{J}{v}\left\{ 1-\exp{\left[ -\frac{\alpha v}{u^2-v^2}\left(x-\frac{\Lambda}{2}\right) \right]} \right\}
    \label{B15}
\end{align}
for each partition shown in Fig.~\ref{fig:fig9}, respectively. Note that we used the relation $F_\mathrm{eff}(c) = F_\mathrm{eff}(x)$ for each partition in Eq.~\eqref{seq:density_formula}. 

In the steady state, Eq.~\eqref{eq:rho&Delta} implies
\begin{align}
	\partial_x J_{\Delta}=-\alpha\Delta.
	\label{seq:J_delta_steady_state}
\end{align}
As discussed above, in case~A, $\rho \equiv \rho_+ + \rho_-$ always stays finite. Since $\rho_+ \ge 0$ and $\rho_- \ge 0$, this also means that $\Delta \equiv \rho_+ - \rho_-$ is finite as well. Then, Eq.~\eqref{seq:J_delta_steady_state} implies the continuity of $J_{\Delta}$ in space.

Using the definitions $J\equiv F_\mathrm{eff}\rho +u\Delta$ and $J_\Delta\equiv F_\mathrm{eff}\Delta +u\rho$, the elimination of $\Delta$ yields
\begin{align} \label{seq:JDelta}
	J_\Delta = \frac{F_\mathrm{eff}}{u}J + \frac{u^2-F_\mathrm{eff}^2}{u}\rho.
\end{align}
Then the continuity of $J_{\Delta}$ at $x=\pm\Lambda/2$ and $x = 0$ leads to
\begin{align}
    -\frac{v}{u}J &+\frac{u^2-v^2}{u}\rho_3(L-\Lambda/2) \nonumber\\
    &=-\frac{\mu F+v}{u}J+\frac{u^2-(\mu F+v)^2}{u}\rho_1(-\Lambda/2),
    \label{seq:continuity_rel1} \\
    -\frac{\mu F+v}{u}J&+\frac{u^2-(\mu F+v)^2}{u}\rho_1(0)\nonumber\\
    &=\frac{\mu F-v}{u}J+\frac{u^2-(\mu F-v)^2}{u}\rho_2(0),
    \label{seq:continuity_rel2} \\
    \frac{\mu F-v}{u}J&+\frac{u^2-(\mu F-v)^2}{u}\rho_2(\Lambda/2)\nonumber\\
    &=-\frac{v}{u}J+\frac{u^2-v^2}{u}\rho_3(\Lambda/2).
    \label{seq:continuity_rel3}
\end{align}
Also, the solution must satisfy the normalization condition
\begin{align}
    \int^0_{-\Lambda/2}{dx\,\rho_1(x)}&+\int^{\Lambda/2}_0{dx\,\rho_2(x)}\nonumber\\
    &+\int^{L-\Lambda/2}_{\Lambda/2}{dx\,\rho_3(x)}= N = \bar{\rho}L.
    \label{seq:normalization}
\end{align}
In case~A, Eqs.~\eqref{seq:continuity_rel1}--\eqref{seq:normalization} fix the boundary conditions of the system. Since we have four unknown parameters $J$, $\rho(\pm l/2^+)$ and $\rho(0^+)$, these four equations completely determine the steady-state density profile and the current $J$. Then, using Eq.~\eqref{B5}, we obtain $F_\mathrm{obj}$.

\subsection{Solution for case~B}

Next we address case~B. As discussed in Appendix.~\ref{subsec:RTP_formula}, $u^2 - F_\mathrm{eff}(x)^2 = 0$ is achieved in the infinitesimal neighborhoods of $x = -\Lambda/2$ and $x = 0$. Let us denote by $c_0$ and $c_0'$ the points at which $F_\mathrm{eff}(x) = -u$ near $x = -\Lambda/2$ and $x = 0$, respectively. We need to check whether $\rho(x)$ diverges to infinity at these points. For this purpose, we express the steady-state current $J$ in terms of the densities of the right-moving and the left-moving RTPs in the neighborhoods of $x = -\Lambda/2$ and $x = 0$. When $F_\mathrm{eff}(x) = -u$, the total effective force on a right-moving particle (including the self-propulsion) disappears. Thus, we can write
\begin{align}
J = -2u\,\rho_-(c_0) = -2u\,\rho_-(c_0'),
\end{align}
which implies that $\rho_-(x)$ stays finite at $x = c_0$ and $x = c_0'$. According to Eq.~\eqref{seq:density_formula}, these are the only points where $\rho(x)$ can possibly diverge. Thus, $\rho_-(x)$ must be finite throughout the system. 

Now it remains to examine the behaviors of $\rho_+(x)$. Near $x = -\Lambda/2$, for positive and infinitesimal $\epsilon$, applying $\int_{-\Lambda/2-\epsilon}^{-\Lambda/2+\epsilon}dx$ to the first identity of Eq.~\eqref{eq:rho+-} in the steady state, we obtain
\begin{align}
(\mu F+v-u)\,\rho_+\!\left(-\frac{\Lambda}{2}+\epsilon \right) &+ (u-v)\,\rho_+\!\left(-\frac{\Lambda}{2}-\epsilon\right)\nonumber\\
&\simeq \frac{\alpha}{2}\int_{-\Lambda/2-\epsilon}^{-\Lambda/2+\epsilon}dx\,\rho_+(x).
\end{align}
Since the left-hand side is bound to be positive, $\rho_+(x)$ must diverge to infinity in the infinitesimal neighborhood of $x = -\Lambda/2$. Since the current
\begin{align} \label{seq:J_rho_pm}
J = \rho_+(x) \left[F_\mathrm{eff}(x)+u\right] + \rho_-(x) \left[F_\mathrm{eff}(x)-u\right]
\end{align}
must be finite, the divergence of $\rho_+(x)$ occurs precisely at $x = c_0$, where $F_\mathrm{eff}(x) + u = 0$. This implies the existence of a delta peak of the RTP density at $x = c_0$. Meanwhile, applying $\int_{-\epsilon}^{\epsilon}dx$ to the first identity of Eq.~\eqref{eq:rho+-} in the steady state, we obtain
\begin{align}
-(\mu F-v+u)\,\rho_+\!\left(\epsilon \right) - (\mu F + v - u)\,\rho_+\!\left(-\epsilon\right)\nonumber\\
= \frac{\alpha}{2}\int_{-\epsilon}^{\epsilon}dx\,\rho_+(x)-\frac{\alpha}{2}\int_{-\epsilon}^\epsilon dx\,\rho_-(x).
\end{align}
Since the left-hand side cannot be greater than zero, the two sides can be equal only if $\rho_+(\pm\epsilon) \sim \epsilon$. Thus, $\rho_+(x)$ converges to zero at $x = c_0'$. As $\rho_-(c_0')$ is finite, this implies that $\rho(c_0')$ is also finite. To sum up, $\rho(x)$ diverges to infinity only at $x = c_0$ in the infinitesimal neighborhood of $x = -\Lambda/2$, and $\rho(x) = \rho_-(x)$ at $x = c_0'$ in the infinitesimal neighborhood of $x = 0$.

Combining the latter observation with Eq.~\eqref{seq:J_rho_pm}, we can show that $J$ directly determines $\rho(0^\pm)$ as follows:
\begin{align}
    \rho(0^{\pm})=\rho_-(0^{\pm})=\frac{J}{F_\mathrm{eff}(0^{\pm})-u}.
    \label{seq:rho_0_PhaseB}
\end{align}
Then, using Eq.~\eqref{seq:density_formula}, we obtain
\begin{align}
    &\rho_1(x)=\rho(0^-)\exp{\left[ -\frac{\alpha(\mu F+v)}{u^2-(\mu F+v)^2}x \right]}\nonumber\\
    &\quad\quad -\frac{J}{\mu F+v}\left\{ 1-\exp{\left[ -\frac{\alpha(\mu F+v)}{u^2-(\mu F+v)^2} x\right]} \right\}
    \label{B24}
\end{align}
for $-\Lambda/2 < x < 0$, where $J$ within $\rho(0^-)$ is the only unknown coefficient. We can similarly express $\rho_2(x)$ and $\rho_3(x)$ in terms of $J$ by applying Eqs.~\eqref{B14}, \eqref{B15}, and \eqref{seq:continuity_rel3}. It should be noted that $\rho_1(x)$, $\rho_2(x)$, and $\rho_3(x)$ are all smooth and finite-valued functions. The delta peak at $x = -\Lambda/2$ must be separately taken into account. Thus, the normalization condition of the RTP density profile can be written as
\begin{align} \label{seq:normalization2}
    \bar{\rho}L=\int^0_{-\Lambda/2}{dx\,\rho_1(x)}&+\int^{\Lambda/2}_0{dx\,\rho_2(x)}\nonumber\\
    &+\int^{L-\Lambda/2}_{\Lambda/2}{dx\,\rho_3(x)}+M,
\end{align}
where $M$ is the magnitude of the delta peak at $x = -\Lambda/2$.

To fully determine the unknown coefficients $J$ and $M$, we revisit Eq.~\eqref{seq:J_delta_steady_state}: $
    \partial_x J_{\Delta}=-\alpha\Delta$. Since the delta peak is entirely due to $\rho_+(x)$, the polarization $\Delta(x)$ also has a delta peak with the same magnitude at $x = -\Lambda/2$. Thus, integrating Eq.~\eqref{seq:J_delta_steady_state} across the infinitesimal interval $[-\Lambda/2-\epsilon,\,-\Lambda/2+\epsilon]$, we obtain
\begin{align}
M = -\frac{1}{\alpha}\left[ J_{\Delta}\!\left(-\frac{\Lambda}{2}+\epsilon\right) - J_{\Delta}\!\left(-\frac{\Lambda}{2}-\epsilon\right)\right].
\end{align}
Using Eq.~\eqref{seq:JDelta}, this can be rewritten as
\begin{align}
    M=-\frac{1}{\alpha}\bigg[ -\frac{\mu F}{u}J+\frac{u^2-(\mu F+v)^2}{u}\rho_1\!\left(-\frac{\Lambda}{2}\right)\nonumber\\
    -\frac{u^2-v^2}{u}\rho_3\!\left(L-\frac{\Lambda}{2}\right) \bigg],
\end{align}
which relates $M$ to $J$. Together with the normalization condition in Eq.~\eqref{seq:normalization2}, this equation fully determines the values of $J$ and $M$. Thus we have fully determined the steady-state RTP density for case~B, and $F_\mathrm{obj}$ can also be derived from $J$ using Eq.~\eqref{B5}.

\section{Small $v$ expansion of the force on the object}
\label{appdx:B}

With $F_\mathrm{obj}$ determined by the procedure described in the preceding appendix, we can expand the expression in terms of small $v$ and single out the leading-order terms for large $L$, getting the linear-order coefficient
\begin{align}
    a_1(f,\lambda,L)
    =&\frac{2}{f} \sinh\left(\frac{f\lambda}{1-f^2}\right)-\frac{\lambda \left(f^4-f^2+2\right)}{\left(1-f^2\right)^2}
\end{align}
and the coefficient of $v^3$,
\begin{align}
    &a_3(f,\lambda,L)
    \simeq-\frac{L}{\Lambda}\bigg[-\frac{\lambda}{3f^2}+\frac{\lambda^3}{6}\frac{(1+f^2)^2}{(1-f^2)^4}\nonumber\\
    &+\frac{\lambda}{3f^2}\cosh\!\left(\frac{f\lambda}{1-f^2}\right)
    -\frac{\lambda^2}{3}\frac{1+f^2}{f(1-f^2)^2}\sinh\!\left(\frac{f\lambda}{1-f^2}\right)\bigg].
\end{align}

\section{Exerted force in the case of the frictional RTP-object interaction}
\label{appdx:C}

In this appendix we aim to provide detailed calculation in obtaining $F_{\mathrm{obj}}$ shown in Eq.~\eqref{eq:fricFobj}. In the steady state, using Eqs.~\eqref{seq:master_eq_friction} and \eqref{seq:fric_obj_force},
\begin{align}
	F_{\text{obj}}&=\frac{1}{\mu}\int^{L}_{0} dx\,(u\Delta-v\rho-J)\nonumber\\
	&=\frac{1}{\mu}\left[ -\frac{u}{\alpha}\int^L_0 dx\,(\partial_x J_\Delta)-v\bar{\rho}L-JL \right]. 
\end{align}
Since the system is periodic, this reproduces the force-current relationship
\begin{align}
	F_{\text{obj}}=-\frac{1}{\mu}(JL+v\bar{\rho}L)
	\label{seq:Fobj}
\end{align}
stated in Eq.~\eqref{B5}.

From Eq.~\eqref{seq:fric_obj_force} we see that no negative drag ($F_\mathrm{obj} > 0$) can occur when there are no right-moving RTPs inside the object ($\rho_+ = 0$). For the negative drag to be possible, there must be right-moving RTPs penetrating the object. As indicated by Eq.~\eqref{seq:fric_motion}, the condition corresponds to $-v+u-f>0$ . Therefore, for the remainder of this appendix we will restrict ourselves to the case $u-f>v>0$, assuming the object to be moving rightward.

In the steady sate, Eq.~\eqref{seq:master_eq_friction} can be rewritten as
\begin{align}
	g'(x)-a(x)\,g(x)+b(x)=0,
	\label{seq:const_eq}
\end{align}
where
\begin{align}
	&g(x)\equiv\frac{[u-f(x)]^2-v^2}{u-f(x)}\rho(x),\\
	&a(x)\equiv-\frac{\alpha v}{[u-f(x)]^2-v^2},\\
	&b(x)\equiv \frac{\alpha-vf'(x)}{u-f(x)}J.
\end{align}
Since the form of the equation is the same as Eq.~\eqref{seq:g}, we can use the same solution as Eq.~\eqref{seq:g_solution}. Denoting by $\rho_1(x)$ and $\rho_2(x)$ the RTP densities inside and outside the object, respectively, we obtain
\begin{align}
	\rho_1(x)=&\rho(-\Lambda/2^+)\exp\left[ -\frac{\alpha v}{(u-f)^2-v^2}\left(x+\frac{\Lambda}{2} \right) \right]\nonumber\\
	&-\frac{J}{v}\left\{ 1-\exp\left[ -\frac{\alpha v}{(u-f)^2-v^2}\left(x+\frac{\Lambda}{2} \right) \right] \right\}, \label{seq:rho1_soln}\\
	\rho_2(x)=&\rho(+\Lambda/2^+)\exp\left[ -\frac{\alpha v}{u^2-v^2}\left(x-\frac{\Lambda}{2} \right) \right]\nonumber\\
	&-\frac{J}{v}\left\{ 1-\exp\left[ -\frac{\alpha v}{u^2-v^2}\left(x-\frac{\Lambda}{2} \right) \right] \right\}. \label{seq:rho2_soln}
\end{align}
Using the definitions of $J$ and $J_\Delta$ in Eq.~\eqref{seq:master_eq_friction}, the elimination of $\Delta$ yields
\begin{align}
	J_\Delta =\frac{[u-f(x)]^2-v^2}{u-f(x)}\rho(x)-\frac{v}{u-f(x)}J.
\end{align}
Within the regime of our interest ($u-f>v>0$), none of the terms in Eq.~\eqref{seq:const_eq} diverge. This implies that $\rho \equiv \rho_+ + \rho_-$ stays finite throughout the system, which in turn implies that $\Delta \equiv \rho_+ - \rho_-$ is also finite. Thus $J_\Delta$, which satisfies $\partial_x J_\Delta = -\alpha \Delta$ in the steady state [see Eq.~\eqref{seq:master_eq_friction}], must be continuous everywhere. The continuity of $J_\Delta$ at $x=\pm\Lambda/2$ leads to
\begin{align}
	&\frac{u^2-v^2}{u}\rho_2(L-\Lambda/2)-\frac{v}{u}J\nonumber\\
	&\quad\quad=\frac{(u-f)^2-v^2}{u-f}\rho_1(-\Lambda/2)-\frac{v}{u-f}J,\label{seq:BC1}
\end{align}
\begin{align}
	&\frac{u^2-v^2}{u}\rho_2(\Lambda/2)-\frac{v}{u}J\nonumber\\
	&\quad\quad=\frac{(u-f)^2-v^2}{u-f}\rho_1(\Lambda/2)-\frac{v}{u-f}J.
\end{align}
The normalization condition for the RTP density,
\begin{align}
	\int^{\Lambda/2}_{-\Lambda/2}dx\,\rho_1(x)+\int^{L-\Lambda/2}_{\Lambda/2}dx\,\rho_2(x)=\bar{\rho}L.
	\label{seq:BC3}
\end{align}
Combining Eqs.~\eqref{seq:rho1_soln}, \eqref{seq:rho2_soln}, and \eqref{seq:BC1}--\eqref{seq:BC3} and taking the limit $L\to\infty$, $J$ is obtained as
\begin{align}
	J\simeq -\bar{\rho}v\left(
	1+\frac{f^2}{\alpha vL}
	\left\{
	1-\exp\left[
	-\frac{\alpha v\Lambda}{(u-f)^2-v^2}
	\right]
	\right\}
	\right).
\end{align}
Applying this result to Eq.~\eqref{seq:Fobj}, we finally obtain Eq.~\eqref{eq:fricFobj}.

\bibliography{symbreak_motility}

\end{document}